\documentstyle[prl,aps,multicol,psfig]{revtex}  
\begin{document}
\newcommand{\be}{\begin{equation}}
\newcommand{\ee}{\end{equation}}
\newcommand{\bq}{\begin{eqnarray}}
\newcommand{\eq}{\end{eqnarray}}
\newcommand{\fat}[1]{\mbox{\boldmath $ #1 $\unboldmath}}
\def\qsqrt{{\sqrt{2} \kern-1.2em ^4}}
\def\CC{{\rm\kern.24em \vrule width.04em height1.46ex depth-.07ex
\kern-.30em C}}
\def\P{{\rm I\kern-.25em P}}
\def\RR{{\rm
         \vrule width.04em height1.58ex depth-.0ex
         \kern-.04em R}}
\def\id{{\rm 1\kern-.22em l}}
\def\ZZ{{\sf Z\kern-.44em Z}}
\def\NN{{\rm I\kern-.20em N}}
\def\up{\uparrow}
\def\dwn{\downarrow}
\def\L{{\cal L}}
\def\E{{\cal E}}
\def\F{{\cal F}}
\def\ga{{\cal G}[sl(2)]}
\def\H{{\cal H}}
\def\U{{\cal U}}
\def\P{{\cal P}}
\def\M{{\cal M}}
\def\W{{\cal W}}
\def\PACS{\par\leavevmode\hbox {\it PACS:\ }}

\title{Exact correlation functions  of the BCS model in the canonical ensemble}
\author{Luigi Amico and Andreas Osterloh}
\address{Dipartimento di Metodologie Fisiche e Chimiche (DMFCI), 
	Universit\'a di Catania, viale A. Doria 6, I-95125 Catania, Italy\\
	Istituto Nazionale per la Fisica della Materia, Unit\'a  di Catania, 
	Italy} 

\maketitle

\begin{abstract}
We evaluate correlation  
functions of the BCS model for finite number of particles.
The integrability 
of the Hamiltonian relates  it with the Gaudin algebra ${\cal G}[sl(2)]$. 
Therefore,  a theorem that Sklyanin proved for the Gaudin model, 
can be applied. Several diagonal and off-diagonal correlators 
are calculated. The finite size scaling behavior of  the pairing 
correlation function is studied.
\end{abstract}
\PACS 02.30.Ik , 74.20.Fg , 03.65.Fd

%%%%%%%%%%%%%%%%%%%%%%%%%%%%%%%%%%%%%%%%%%%%%%%%%%%%%%%%%%%%%%%%%%%

\begin{multicols}{2}
{\it Introduction.}
The BCS model expresses  the competition between the kinetic energy and the 
tendency of  Cooper pairs to condense~\cite{BCS}. 
This idea has been successfully  employed  to a large variety of different physical context 
such as superconductivity~\cite{TINKHAM}, 
nuclear physics~\cite{IACHELLO94},  
QCD~\cite{RISCHKE} and astrophysics\cite{ASTRO}. 
The BCS Hamiltonian is 
\begin{eqnarray}
H  = \sum_{j=1 \atop \sigma= \up, \dwn}^\Omega 
\varepsilon_{j} n_{j \sigma}
  -g  \sum_{i, j=1}^\Omega   c_{i\up }^\dagger c_{i \dwn}^\dagger 
c^{}_{j \dwn} c^{}_{j \up} .
\label{pairing}
\end{eqnarray}
$g$ is the pairing coupling constant; 
the quantum numbers $j\in \{1\dots \Omega\}$  label the 
single particle energy levels
$\epsilon_j$ which are doubly degenerate since  
 $\sigma \in \{\up, \dwn\}$  labels 
time reversed electron states; 
$c_{j,\sigma}$ and $n_{j,\sigma}:=c_{j \sigma}^\dagger c_{j \sigma}$ 
are annihilation and number operators. 
\\
In the grand canonical ensemble, the mean field approximation of the 
BCS model has been successfully applied~\cite{TINKHAM,BOGOLIUBOV}. 
However, 
the grand canonical analysis is not appropriate
in many important physical situations (noticeably in nuclei).
Recent experiments on small metallic grains~\cite{BRT} 
evidenced this limitation also 
in condensed matter~\cite{BRAUN,NOTE}. Indeed,
the typically very low capacitance 
of  metals of nanoscale size  {\it fixes} the number of particles 
in the grain. This has constituted a conceptual challenge\cite{ANDERSON},
since physical quantities must be studied in the {\it canonical ensemble}. 
In particular, the system cannot undergo the
superconducting phase transition, but a cross-over regime 
dominated by superconducting fluctuations\cite{MASTELLONE}.  
The quantity playing  the role of the 
``order parameter'' is  
the  pairing correlation function 
$u_{ij}:=\langle  
c_{i,\dwn} c_{i,\up} 
c^\dagger_{j,\up} c^\dagger_{j,\dwn}
\rangle$.
Canonical pairing fluctuations were studied in the recent 
literature  employing numerical and analytical 
techniques~\cite{MASTELLONE,OTHERS}. 
Since the system is characterized by 
strong quantum fluctuations, its  physical behavior 
is very sensitive to the approximations employed and therefore
 exact results play an important role.
The BCS model is Bethe 
Ansatz (BA) solvable~\cite{RICHARDSON}   (see also Ref.~\cite{GAUDIN}) and   
more recently found to be integrable~\cite{CAMBIAGGIO}.
The integrals of motion obtained in Ref.~\cite{CAMBIAGGIO} 
can be  also generated within the Quantum Inverse Scattering  
(QIS) scenario through the quasi classical limit of the $R$--matrix 
of disordered six vertex models~\cite{SKLYANIN-GAUDIN,SIERRA,AMICO}.  
By these integrals, the 
model becomes connected with the Gaudin Hamiltonians. 
%which are 
%the BCS constants of motions 
%in the limit  $g\rightarrow \infty$.  
The BA solution played a central 
role for theoretical predictions in particular  of 
the low temperature behavior of 
thermodynamic quantities of 
grains~\cite{DILORENZO,DUKELSKY,VONDELFT}. 
Indeed, for systems with a fixed number of particles 
(metallic grains but also nuclei) the BA solution is particularly useful
and feasible even beyond the thermodynamic limit 
(which is rare in condensed matter) due to its simplicity.
Using the  exact eigenstates  of the 
Hamiltonian, the {\it diagonal} pairing correlation 
function was obtained  by Richardson although it was not    
evaluated explicitly~\cite{RICHARDSON-CORRELATION}.
The aim of the present work is to fill these gaps; 
we evaluate exactly {\it diagonal and off-diagonal} correlation  
functions of the BCS model for finite number of particles.

In the realm of exactly solvable models Correlation Functions (CFs) 
are studied resorting various approaches.
For large systems (in the thermodynamic limit)
they have been  extracted 
from analytic properties 
of form factors~\cite{JIMBO}.
For spin models,  CFs  have been  
studied by Korepin using the operator algebra from
the QIS method~\cite{KOREPIN-BOOK} to 
calculate the 
scalar products of Bethe states. 
In a recent paper~\cite{SKLYANIN-CORRELATORS} Sklyanin suggested 
how the combinatoric complications  involved in this calculations can be 
overcome resorting  the Generating 
Function (GF) technique. 
He applied it to the $sl(2)$ Gaudin model~\cite{GAUDIN,SKLYANIN-GAUDIN}.
The key role in his approach is played by an analog 
of the Gauss decomposition or Baker--Campbell--Hausdorff (BCH) formula 
for elements of the $SL(2)$ loop group associated to the 
Gaudin algebra ${\cal G}[sl(2)]$. 
In the present work, the latter approach is pursued.
We exploit  the common algebraic root of the Gaudin 
and BCS models
to extend the Sklyanin theorem to the BCS model. 
The  GF of exact 
CFs for the BCS model is obtained.
The CFs are suitable residues of the GF (see 
Eqs.~(\ref{charge-residue}), (\ref{limit-pairing})). 
The $M$-point charge and pairing correlations  
\vspace*{-1mm}
\begin{equation}
\pi_0({\cal E},M,\F):=
\langle \E|\,
\prod_{k=1}^M 
( n_{j_k,\up}+  n_{j_k,\dwn}-1)/2 \,|\F \rangle
\label{n-correlators}
\end{equation}
\begin{equation}
u_{ij}({\cal E},\F):= 
\langle \E| 
c^{}_{i,\dwn} c^{}_{i,\up} 
c^\dagger_{j,\up} c^\dagger_{j,\dwn}
 \, |\F \rangle
\label{pairing-correlator}
\end{equation}
are computed exactly in the canonical ensemble
(see Eqs.~(\ref{n-charge-correlators}), ~(\ref{problem})
and Figs.~(\ref{scaling}), (\ref{corr-fig})).
The vectors  $\langle \E|$, 
$|\F \rangle$ are exact $N$-pair eigenstates of (\ref{pairing})
(see Eqs.~(\ref{gaudin-operators}), (\ref{BA-state})). 
%Off-diagonal correlations correspond to $\langle \E|\neq \langle \F| $,
%diagonal to $\langle \E|\equiv \langle \F|$. 

The present paper is laid out  as follows. First we  fix the algebraic aspects 
we  employ in the paper. This motivates the extension  of 
the Sklyanin theorem to the BCS model. Then we evaluate the charge and the 
pairing CFs. Finally we will draw our conclusions.

\bigskip

%%%%%%%%%%%%%%%%%%%%%%%%%%%%%%%%%%%%%%%%%%%%%%%%%%%%%%%%%%%%

\vspace*{-0.2cm}

{\it The BCS model and the Gaudin algebra}.
We first sketch the connection of the BCS model with the 
$sl(2)$-Gaudin model. For this goal we 
introduce the fundamental  realization of   
$su(2)\simeq sl(2)$ in terms of electron pairs is 
$
S^-_j := c_{j,\dwn} c_{j,\up}, \;\; S^+_j:=(S^-_j)^\dagger= 
c^\dagger_{j,\up} c^\dagger_{j,\dwn} , \;\; 
S^z_j := (c^\dagger_{j,\up} c_{j,\up}+ c^\dagger_{j,\dwn} c_{j,\dwn}-1)/2 
$. 
The  $sl(2)$  ``lowest''  weight module  is generated by the vacuum
vector $|0\rangle_j  $, 
$ 
S^-_j |0\rangle_j=0\;, \quad S^z_j |0\rangle_j=s_j |0\rangle_j
$
where $s_j$ is the ``lowest''  weight ($s_j=-1/2$ for spin $1/2$, 
which is the case of interest here~\cite{SKLYANIN-NOTATION}). 
The quadratic Casimir operator is:
$
S_j^2:= (S^z_j)^2+
{{1}\over{2}}\left(S^+_j S^-_j + S^-_j S^+_j\right ) 
$, 
$S_j^2|0\rangle_j=s_j (s_j-1) |0\rangle_j$.
The bilinear combinations $S^+_j S^-_j$ and  
$S^-_j S^+_j$ can be expressed in terms of Casimir and Cartan 
operators 
\begin{equation}
S^\pm_j S^\mp_j=S^2_j-(S^z_j)^2 \pm S^z_j\;. 
\label{pairing-operators}
\end{equation}
The integrals of motion $\tau_l$ of the BCS model 
are  $\tau_l = S^z_l/g -  \; \Xi_l$, that is:  
$
[H, \tau_j] = 
 [\tau_j, \tau_l] = 0$ for $j, l =1, \dots, \Omega$~\cite{CAMBIAGGIO}. 
By these integrals of motion, the 
model becomes connected with the Gaudin Hamiltonians 
$
\Xi_j:= \sum_{l\neq j}^\Omega 
{
{\fat{S }}_j \cdot {\fat{S}}_l }/{( \varepsilon_j- \varepsilon_l)}  
$,
where ${\fat S}_j:= (S^x_j, S^y_j,S^z_j)$; 
$S^\pm_j=1/\sqrt{2}(S^x_j\pm i S^y_j)$) are spin vectors.   
The Hamiltonian~(\ref{pairing}) can be 
expressed in terms of $\tau_j$:
$
 H = g \sum_{j=1}^\Omega 2 \varepsilon_j \tau_j +  g^3\sum_{j,l=1}^\Omega \tau_j  \tau_l  + const.
$
However, the relation is deeper and the integrability of the BCS model
is founded in its connection with 
the infinite dimensional Gaudin algebra 
${\cal G}[sl(2)]$ that is constructed  from $sl(2)$ as
\begin{equation}
S^\pm(u) := \sum_{j=1}^\Omega \frac{S^\pm_j}{ u-2 \varepsilon_j } 
\; , \; 
S^z(u) := \sum_{j=1}^\Omega \frac{S^z_j}{u-2 \varepsilon_j} \; .
\label{gaudin-operators}
\end{equation}
The lowest weight module of ${\cal G}[sl(2)]$ is generated by the vacuum 
$|0\rangle\equiv \otimes_{j=1}^\Omega |0\rangle_{j}$:
$
S^-(u) |0\rangle=0 
\; , \quad S^z(u) |0\rangle
=s(u) |0\rangle \; ,
$
where 
$s(u):=\sum_{j=1}^\Omega {s_j}/{(u-2 \varepsilon_j)} $ 
is the lowest  weight  of ${\cal G}[sl(2)]$.
The mutual commutativity of $\tau_j$ descends from the 
relation between 
$\tau(u):= \sum_{j=1}^\Omega {{\tau_j}/{(u-2 \varepsilon_j)}}$ and 
invariants (trace and quantum determinant\cite{SKLYANIN-GAUDIN}) 
of $\ga$. It can be written as
$\tau (u) =t(u) + s^{[2]}(u)$\cite{taunote}
where $s^{[2]}(u):=\sum_{j=1}^\Omega {{s_j }/{(u-2 \varepsilon_j)^2}} 
$;   $t(u):=-2 S(u)+S^z(u)/{g} $ and 
$
S(u):=S^z(u)S^z(u) + {{1}\over{2}}\left( S^+(u)S^-(u)+S^-(u)S^+(u) \right)
$. 
The  property  $[t(u),t(v)]=0$ is the ultimate reason for the integrability
of the BCS model. 
Accordingly, the exact eigenstates of both the BCS  model~\cite{RICHARDSON,GAUDIN} 
and the set of operators $\tau_j$\cite{SKLYANIN-GAUDIN} are constructed 
from $\ga$ generators
\begin{eqnarray} 
\label{BA-state} 
|\E \rangle_N = \prod_{\alpha = 1}^N &&S^+(e_\alpha) |{\rm 0} \rangle \quad , 
\end{eqnarray}
$ H| \E\rangle_N 
= E |\E \rangle_N  
$; the energy $\; E = \sum_{\alpha =   1}^{N} e_\alpha$ is 
given in terms of the set  ${\cal E}$ of the  
spectral parameters $e_\alpha$ 
satisfying  the algebraic Richardson-Sherman (RS) equations
\begin{equation}
\label{re}
s(e_\alpha) =\frac{1}{2g } + \sum_{\beta=1 \atop \beta\neq \alpha}^{N} \frac{1}{ e_\beta- e_\alpha} 
\; , \quad \alpha = 1, \dots , N  \; .
\end{equation}
We note that RS equations~(\ref{re}) are intimately related to  
the algebraic structure of ${\cal G}[sl(2)]$ 
since they act as constraints on  the lowest  weight $s(e_\alpha)$. 
Thus,  the difference between the BCS and Gaudin model 
results in a different constraint imposed on the lowest weight 
vector of ${\cal G}[sl(2)]$ which leads to different sets ${\cal E}$, 
${\cal E}'$ ($\E' $ is spanned by the solutions of ~(\ref{re}) when $g\to \infty$).
We will use this fact to extend the Sklyanin theorem 
to the BCS model.  

\bigskip
\vspace*{-0.2cm}

{\it Generating functions}.
CFs of elements in $\ga$ can be expressed in terms 
of the following GF
\begin{eqnarray}
C({\cal E}, {\cal H}, {\cal F}):=
\langle \F |\prod_{h \in {\cal H}}  S^z(h) |\E \rangle  
\label{general-correlators}
\end{eqnarray}
where  $\langle \F |:=\langle 0| \prod_{f_\beta \in {\cal F}}S^-(f_\beta)$ 
and the sets $\E, \F \subset \CC \setminus \E_0$ are 
(in general distinct) sets of solutions of the  RS equations~(\ref{re});
$\E_0:=\{2 \varepsilon_j, j=1\dots\Omega\}$; 
$\H \subset \CC \setminus (\E\cup\F\cup \E_0)$.  
The order of the correlation is the cardinality of $\H$: $|{\cal H}|$;
$|{\cal E}|$ and $|{\cal F}|$ are  fixed 
by the number of pairs $N$.
For instance, the one and two point CFs 
correspond to $|{\cal H}|=1$ and $|{\cal H}|=2$ respectively.    

Now we present the Sklyanin theorem for GF of $sl(2)$ 
Gaudin model and apply it to the BCS model.
Therefore we need the notation of 
the set of {\it coordinated partitions} 
${\cal P}= \{ P_l : l\in 1 \dots  |{\cal P}|\}$ 
of the sets ${\cal E}, \F, {\cal H}$ 
(see Ref.~\cite{SKLYANIN-CORRELATORS}):
the partition  $P\in \P$ is a set of triplets
$\{T_1\dots T_{|P|} \}$; the triplet $T\in P$ is  
$T=(\E_T,  \F_T, 
{\cal H}_T) $, where 
$\emptyset \neq {\cal E}_T \subset \E$, 
$\emptyset \neq {\cal F}_T \subset \F$ 
and ${\cal H}_T \subset {\cal H}$ 
such that $|\E_T|= |\F_T| >0, 
\quad |\H_T|\geq 0$. 
%Then $P$ is a coordinated partition $iff$ 
% for  $T\neq T^{'}$ 
%\begin{eqnarray}
%&& \E_T\cap\E_{T^{'}}=\emptyset, 
%\quad \F_T\cap\F_{T^{'}}=\emptyset ,
%\quad \H_T\cap\H_{T^{'}}=\emptyset \; , \\
%&& \bigcup_{T\in P}\E_T=\E
%\quad \bigcup_{T\in P}\F_T=\F , 
%\qquad H_P:=\bigcup_{T\in P}\H_T\subset\H \;. \nonumber  
%\end{eqnarray}
%It follows that:
%$ \sum_{T\in P} 
%|\E_T|=
%  \sum_{T\in P} |\F_T|=N, \quad
%  \sum_{T\in P}|\H_T|\leq|\H|=M $.
\\
The GF has  been evaluated  for the $sl(2)$ Gaudin 
model exploiting the BCH formula
for 
the  $SL(2)$ loop group generated by 
$S^-_{\phi(x)}:=\sum_{f\in \F}\phi_f S^-(f)$, $S^z_{\eta(x)}:=\sum_{h\in \H}
\eta_h S^z(h)$, 
$S^+_{\psi(x)}:=\sum_{e\in \E}\psi_e S^+(e)$ where $\{S^z(u),S^\pm(u)\}\in \ga$ and  
$\phi(x),\, \eta(x),\, \psi(x)$ are meromorphic functions for  $x\in \CC$
with residues $\phi_f,\, \eta_h,\, \psi_e$ respectively~\cite{SKLYANIN-CORRELATORS}. 
This formula allows 
to reorder  the products between loop group elements 
in ~(\ref{general-correlators}):
$\langle\exp{S^-_{\phi(x)}}\exp{S^z_{\eta(x)}}\exp{S^+_{\psi(x)}} \rangle
\equiv\langle\exp{S^+_{\tilde{\psi}(x)}}\exp{S^z_{\tilde{\eta}(x)}}
\exp{S^-_{\tilde{\phi}(x)}} \rangle=
\langle\exp{S^z_{\tilde{\eta}(x)}}\rangle$.  
Sklyanin proved the following theorem~\cite{SKLYANIN-NOTATION}.

{\bf Theorem}.
$C(\E,\H,\F)$
{\it is given by the formula
\begin{eqnarray}
\label{sklyanin-theorem}
\lefteqn{
C(\E,\H,\F)=(-1)^N \times} \\
&&\times \sum_{\P}\left(\prod_{T\in P}
n_{T}( |\E_T|)^{|\H_T|} 
S(\W_T\cup\H_T)\right) \prod_{h\in{\overline\H}_P} s(h) \nonumber 
\end{eqnarray}
where 
$S({\cal L})=1/2\pi i {\int}_\Gamma \; s(z)
\prod_{y\in \L} (z-y)^{-1}   dz$
%The exact evaluation of expression~(\ref{general-correlators}) 
%can be written in terms of 
%the quantity
%\begin{equation}
%S({\cal L})={{1}\over{2\pi i}} {\int}_\Gamma \; \frac{s(z)}
%{\prod_{y\in \L} (z-y)}   dz
%\label{S-expression2}
%\end{equation}
~\cite{ANALYTIC};
$ n_T:=-2 |\E_T|!\,(|\E_T|- 1)!$,
$\W_T:=\E_T\cup\F_T$,  
and ${\overline \H}_P:=
\H \setminus \bigcup_{T\in P} \H_T$. 
$C(\E,\H,\F)$ is a polynomial 
in $S$ with integer coefficients.} 
\\
Expression (\ref{sklyanin-theorem}) 
depends only on the sets 
$\W:=\E\cup\F$ and $\H$~\cite{SKLYANIN-CORRELATORS,KOREPIN-GAUDIN}; 
for the Gaudin model $\W$ is a set of 
solutions of ~(\ref{re}) for $g \to \infty$;
for the BCS model $\W$ is a set of solutions of the  RS~(\ref{re}) 
for generic $g$.  
The scalar products of Bethe states (and their norms)
are a  corollary 
of the Sklyanin theorem~(\ref{sklyanin-theorem}) for $\H=\emptyset$:  
$\langle \E | \F \rangle=C(\E,\emptyset,\F)$.
Its concent with the determinant 
formulas\cite{GAUDIN,RICHARDSON-CORRELATION} has been elucidated in 
Refs.\cite{SKLYANIN-CORRELATORS,KOREPIN-GAUDIN}.
\\
We point out that the GF~(\ref{general-correlators}) 
has simple poles in the set $\E_0$. This will play 
a key role in the following.

%%%%%%%%%%%%%%%%%%%%%%%%%%%%%%%%%%%%%%%%%%%%%%%%%%%%%%%%%%%%%

\bigskip
\vspace*{-0.2cm}
{\it Correlation functions}.
The charge and the pairing CFs
are  matrix elements of $su(2)$ Lie algebra (instead of elements of $\ga$) 
using vector states of $\ga$. 
The projection from the $sl(2)$ loop algebra on its Lie algebra 
is performed by taking the residue of $C(\E,\H,\F)$ in the poles 
$h_l=2\varepsilon_{j_l}$ 
for $h_l\in \H$, $l \in \{ 1 \dots M\}$.
The charge CFs~(\ref{n-correlators}) are 
\begin{equation}
\pi_0({\cal E}, M,\F)= 
\lim_{\H \to \E_0 } (\H-\E_0)C(\E,\H,\F) 
\label{charge-residue}
\end{equation}
where $\H \to \E_0$ and $\H-\E_0$ mean $ h_l \to 2 \varepsilon_{j_l}$ 
$\forall l$ and 
$\prod_l (h_l -2 \varepsilon_{j_l})$ respectively. 
Using (\ref{sklyanin-theorem}) yields 
\begin{eqnarray}
\label{n-charge-correlators}
&&\pi_0({\cal E}, M,\F) =(-1)^N \prod_l^{M}s_{j_l} \times  \\
&&\sum_{P \in \P_1}  \left ( \prod_{T\in T_0} n_{T} 
S(\W_T)  \right )
\left (\prod_{T\in T_1 \atop \H_T = h_T}  
  \frac{ n_{T}|\E_T| }
{\prod\limits_{y\in \W_T} (h_T -y)}\right) 
\nonumber
\end{eqnarray}
where  $\P_k\equiv \{P\in \P : \max\limits_{T\in P} |H_T| = k \}$;
$T_k\equiv \{T\in P : |H_T| = k \}$. The quantity $S(\W_T)$ is
\begin{eqnarray}
&&S(\W_T)= \sum_{e \in \W_T} 
\frac{s(e)}{\prod\limits_{x\in \W_T \atop x \neq e} (e -x)} - \\ 
&&\sum_{d \in \W_T}\left (\frac{s(d)}{\prod\limits_{x\in \W_T \atop x \neq d} (d -x)}
\sum_{y \in \W_T \atop y \neq d }\frac{1}{(d -y)}+ \frac{s^{[2]}(d)}
{\prod\limits_{x\in \W_T \atop x \neq d} (d -x)}\right ) \nonumber    
\end{eqnarray}
where $e$ and $d$ are elements 
appearing singly and doubly in $\W_T$ respectively. 
The pairing CF~(\ref{pairing-correlator}) can be  
extracted  from  $C(\tilde{\E},\emptyset,\tilde{\F})$ where   
the vectors in ~(\ref{general-correlators}) are 
$\langle \tilde{\E}|:= \langle \E| S^-(z_1)$ and 
$|\tilde{\F}\rangle :=S^+(z_2) |\F\rangle $.
Then $u_{lm}(\E,\F)$ is   
\begin{eqnarray}
u_{lm}(\E,\F) =\lim_{\scriptstyle z_1\to 2 \varepsilon_l 
\atop \scriptstyle z_2\to 2 \varepsilon_m}( z_1- 2 \varepsilon_l) 
( z_2- 2 \varepsilon_m)  C(\tilde{\E},\emptyset ,\tilde{\F})
\label{limit-pairing}
\end{eqnarray}
$C(\tilde{\E},\emptyset ,\tilde{\F})$ can be calculated using the Sklyanin theorem.
For $l\neq m$ formula (\ref{limit-pairing})  gives
\begin{eqnarray}
\label{problem}
&&u_{lm}({\cal E},\F)  =(-1)^{N+1} \\ 
&&\sum_{P \in \tilde{\P}_1}  \left (\prod_{T\in \tilde{T}_0} n_{T} 
S(\W_T) \right ) 
\left (\prod_{T\in \tilde{T}_1} 
  \frac{n_{T}  s_{l_T}}
{\prod\limits_{y\in \W_T} 
(2 \varepsilon_{l_T} -y) }\right) 
\nonumber
\end{eqnarray}
where $Z:=\{z_1,z_2\}$;  
$\tilde{\P}_k\equiv \{P\in \P : \max\limits_{T\in P} |Z_T| = k \}$;
$\tilde{T}_k\equiv \{T\in P : |Z_T| = k \}$. 
The pairing CF for 
$l\equiv m$ can be achieved by 
a variation  of the  procedure 
depicted above. Namely by 
$\lim_{\scriptstyle z_1\to 2 \varepsilon_l 
\atop \scriptstyle z_2\to 2 \varepsilon_l}( z_1- 2 \varepsilon_l) 
( z_2- 2 \varepsilon_l)  C(\tilde{\E},\emptyset ,\tilde{\F})$.
But in the present case ($s_j=-1/2\, \forall j$) 
it is more convenient employing the formula~(\ref{pairing-operators}) 
which  simplifies in
$S^\pm_j S^\mp_j=1/2 \pm S^z_j$. 
Then  the CF 
$\Psi:=\sum_j u_j v_j:=\sum_j 
\sqrt{\langle S^-_j S^+_j\rangle \langle S^+_j S^-_j\rangle }=\sum_j 
\sqrt{1/4-\langle S^z_j\rangle^2}$
can be calculated evaluating 
$\langle S^z_j \rangle$ by the formula~(\ref{n-charge-correlators}) 
for $M=1$. 
\vspace*{-10mm}
\begin{center}
\parbox[t]{8.5cm}{
\begin{figure}
{\psfig{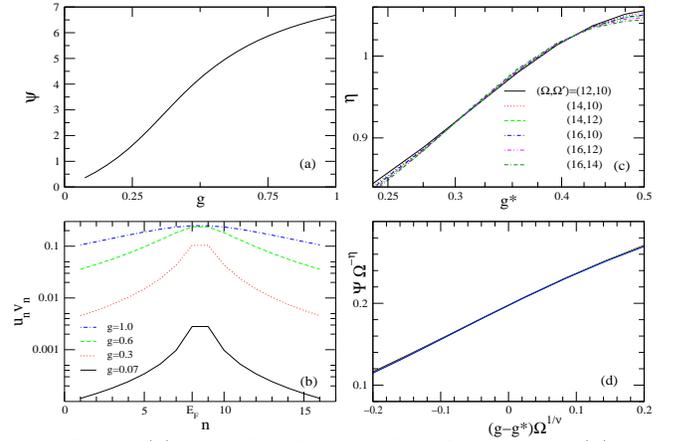}}
\caption{ (a) $\Psi$ as function of $g$
for $16$ electrons. 
 (b) The Cooper pairs probability density vs their energy 
$\varepsilon_n=n$ (8 pairs). 
 (c) The figure shows two  crossing points: the first
is in agreement with Ref.~[11]; the second is at 
$g^*=0.417$, $\eta=1.028$. (d) The data collapse  in 
$g^*$ for $1/\nu=0.15$.}
\label{scaling}
\end{figure}}
\end{center}
\vspace{-0.3cm}
In the plots, only systems with half-filling ($\Omega=2N$) are 
calculated with Eqs. (\ref{n-charge-correlators}) 
and (\ref{problem})~\cite{NOTCRUCIAL}.
In order to extract  the  finite size scaling 
of the 'order parameter' $\Psi$
 we plotted the function 
$\log(\Psi_\Omega(g)/\Psi_{\Omega'}(g))/\log(\Omega/\Omega')=:
\eta(\Omega,\Omega';g)$ for distinct 
$\Omega,\Omega'\in\{10,12,14,16\}$. At a scaling point $g^*$, all
such curves cross in a single point: 
$\eta(\Omega,\Omega';g^*)\equiv\eta(g^*)$. 
In addition to the crossing point found in Ref.~\cite{MASTELLONE}
we detected a second one (Fig.~\ref{scaling} (c), (d)).
Hence, the analysis done in Ref.~\cite{MASTELLONE} seems to require
further investigations to characterize 
the superconducting instability in small grains. 
In Fig.~\ref{corr-fig} we plot some off diagonal CFs.  
They can be useful to clarify the effects of 
BCS pairing correlations in tunneling experiments (notice the
non-monotonic behavior in $g$ of Fig.~\ref{corr-fig} (b)).

\bigskip
\vspace*{-0.2cm}

{\em Conclusions}. We have evaluated exactly charge and  pairing 
CFs of the BCS model in the canonical ensemble.
They are the main results of this paper (Eqs.~(\ref{n-charge-correlators}),
(\ref{problem}), Figs.~(\ref{scaling}), (\ref{corr-fig})). 
%For the first time 
For this goal the algebraic connection of the BCS model with the 
Gaudin model is crucial and it is emphasized and exploited for the first time.
General charge and pairing correlations are obtained as certain 
residues of the GF (a ``correlator'' within $\ga$) 
via a theorem for CFs of $su(2)$ operators in the lowest weight module
of $\ga$\cite{SKLYANIN-CORRELATORS}.
The limitation of the numerics is
the vastly increasing number of partitions, which
depends on the number of pairs $|\E|=N$ and the order of the 
CF $|\H|$. We want to emphasize that it does not
depend on the dimension of the Hilbert space $\Omega$.
\vspace*{-5mm}
\begin{center}
\parbox[t]{8.5cm}{
\begin{figure}
{\psfig{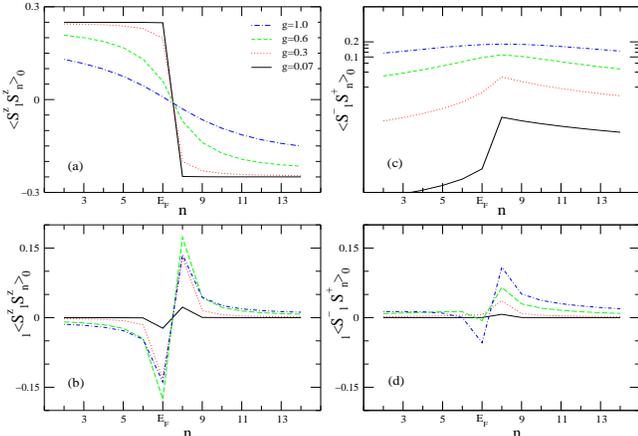}}
\caption{Correlators are plotted as function of $\varepsilon_n=n$ for a 
14 electron system.
(a) the charge correlation
$\pi_0(\E_g,\{1,n\},\E_g)$, where $\E_g$ represents the ground state.
(b) $\pi_0(\E_g,\{1,n\},\E_e)$, where $\E_e$ represents 
the first excited state.
(c) The ground state pairing correlation $u_{1,n}(\E_g,\E_g)$. 
(d) The off-diagonal pairing correlation $u_{1,n}(\E_g,\E_e)$.}
\label{corr-fig}
\end{figure}}
\end{center}
\vspace{-0.3cm}
%, once
%the solutions of the RS equations are obtained.
We roughly estimated the complexity involved in the evaluation of our 
formulas and compared it with the complexity 
of the corresponding diagonal quantities calculated in  
Ref.~\cite{RICHARDSON-CORRELATION}.
We found that for diagonal form factors it is favorable to use the expression
in Ref.~\cite{RICHARDSON-CORRELATION}, whereas
the complexity of the formula 
presented here becomes much lower for the $2$-point functions. 
%A numerical confrontation would be desirable to decide definitely which
%formula should be used for what quantities. Up to now, the numerical
%evaluation of CFs presented here is the only one 
%existing in the literature.
Our results might have 
immediate physical relevance for metallic grains, 
invoking the universality of their physical properties for very small 
systems already, as shown in Ref.\cite{MASTELLONE}.
In addition our formulas can be the starting point 
for refinements  entering the RS equations
in them following the scheme developed in~\cite{KOREPIN-BOOK}; 
%Eqs.~(\ref{n-charge-correlators}), (\ref{problem}) 
they pave the way 
towards asymptotics of the CFs and thermal Green functions.
We finally point out that our results apply to arbitrary $s_j$ 
(i.e. any degeneracy of the single particle levels);
we consider $s_j=-1/2$ in the plots. \\
%\bigskip
{\em Acknowledgments}: 
We acknowledge  A. Mastellone for useful discussions and 
important assistance in numerics.
We thank  R. Fazio and G. Falci for constant support, fruitful  discussion and 
for a critical reading of the manuscript.
We acknowledge D. Boese, F. Castiglione, G. Giaquinta, and
F. Taddei for useful discussions.

%%%%%%%%%%%%%%%%%%%%%%%%%%%%%%%%%%%%%%%%%%%%%%%%%%%%%%%%%
%%%%%%%%%%%%%%%%%%%%%%%%%%%%%%%%%%%%%%%%%%%%%%%%%%%%%%%%

%\vspace*{-1mm}

%%%%%%%%%%%%%%%%%%%%%%%%%%%%%%%%%%%%%%%%%%%%%%%%%%%%%%%%%%%%%%%%%%%%%%%%%%%%%%%
%%%%%%%%%%%%%%%%%%%%%%%%%%%%%%%%%%%%%%%%%%%%%%%%%%%%%%%%%%%%%%%%%%%%%%%%%%%%%%

\end{multicols}

\begin{thebibliography}{100}
\bibitem{SKLYANIN-NOTATION} In the Ref.~\cite{SKLYANIN-CORRELATORS} 
	a different notation  was adopted  in that 
	the highest weight vector $s_j=1/2$  as vacuum was used.
\bibitem{taunote} The $\tau_j$ are residues of 
	$\tau(u):= \sum_{j=1}^\Omega {{\tau_j}/{(u-2 \varepsilon_j)}}$ 
	in $u=2 \varepsilon_l$; note the analogy with 
	Eq.~(\ref{gaudin-operators}).
\bibitem{ANALYTIC} $s(z)$ is assumed to be analytic inside the contour 
	$\Gamma$ encircling ${\cal L}\subset \CC$;
	therefore, $\E_0$ must be outside $\Gamma$. $\E$, $\F$, and $\H$ 
	are to be inside $\Gamma$;
	$\Gamma$ placed in the open set $\{\E_0,\E,\F,\H$ 
	pairwise disjoint$\}$.
	Afterwards, limits of $\E,\F,\H$ towards $\E_0$ can be taken
	for the analytic GF.
\end{thebibliography}
\end{document}